\documentclass[sigconf]{acmart}

\usepackage[indentfirst=false,font={itshape},begintext=``,endtext='']{quoting}
\usepackage{subcaption}
\usepackage{hyperref}
\usepackage[most]{tcolorbox}
\newtcolorbox{summarybox}{
    enhanced,
    breakable,
    sharp corners,
    boxrule=0.3pt,
    colback=white,
    colframe=black!40,
    coltitle=black,
    fonttitle=\bfseries,
    toptitle=2mm,
    bottomtitle=2mm,
    colbacktitle=white,
}

\usepackage{longtable,booktabs,array}
\usepackage{calc} 
\usepackage{etoolbox}

\patchcmd\longtable{\par}{\if@noskipsec\mbox{}\fi\par}{}{}

\AtBeginDocument{%
  \providecommand\BibTeX{{%
    \normalfont B\kern-0.5em{\scshape i\kern-0.25em b}\kern-0.8em\TeX}}}

\acmConference[GAS '22]{6th International ICSE Workshop on Games and Software Engineering:  Engineering fun, inspiration, and motivation}{June 03--05, 2018}{Pittsburgh, PA, USA}

\makeatletter

\usepackage{amssymb}
\newcommand{\ygg@basicalert}[2]{\textcolor{red}{\fbox{\bfseries\sffamily\scriptsize#1}{\sf\small$\blacktriangleright$\textit{#2}$\blacktriangleleft$}}}
\newcommand{\YANN}[1]{\ygg@basicalert{YANN}{#1}}

\begin{document}

\title{Towards Automated Video Game Testing: Still a Long Way to Go}

\author{Cristiano Politowski}
\email{c_polito@encs.concordia.ca}
\orcid{0000-0002-0206-1056}
\author{Yann-Ga\"el Gu\'{e}h\'{e}neuc}
\email{yann-gael.gueheneuc@concordia.ca}
\orcid{0000-0002-4361-2563}
\affiliation{%
  \institution{Concordia University}
  \city{Montreal}
  \state{Quebec}
  \country{Canada}
}

\author{Fabio Petrillo}
\email{fabio@petrillo.com}
\orcid{0000-0002-8355-1494}
\affiliation{%
  \institution{Université du Québec à Chicoutimi}
  \city{Chicoutimi}
  \state{Quebec}
  \country{Canada}
}

\begin{abstract}
As the complexity and scope of game development increases, playtesting remains an essential activity to ensure the quality of video games. Yet, the manual, ad-hoc nature of playtesting gives space to improvements in the process. In this study, we investigate gaps between academic solutions in the literature for automated video game testing and the needs of video game developers in the industry. We performed a literature review on video game automated testing and applied an online survey with video game developers. The literature results show a rise in research topics related to automated video game testing. The survey results show that game developers are skeptical about using automated agents to test games. We conclude that there is a need for new testing approaches that did not disrupt the developer workflow. As for the researchers, the focus should be on the testing goal and testing oracle.
\end{abstract}


\maketitle

\section{Introduction}
\label{sec:intro}

Recently, a former Playstation executive explained that the cost of AAA video games is doubling with each new console generation. He expects that every new big PS5 game will cost 200 million dollars, at least. This skews the game industry towards safer alternatives to new games, like sequels\footnote{\url{https://www.bloomberg.com/news/newsletters/2021-09-03/ex-playstation-chief-mulls-future-of-gaming-and-his-new-job}}.
    
New games are challenging because the complexity of game development increases with their scopes. Bigger games need more work, time, people, and funding. Also, users (players) expect a bigger and better game every new release. Yet, to achieve success in the saturated video game market, games must also be of quality. Thus, to succeed commercially, game developers must reduce expenses while keeping the high-quality levels in their games.

Playtesting is one of the activities to assess a game's quality \cite{albaghajatiVideoGameAutomated2020}. It consists of manually playing the game while assessing its details; for example, checking whether the game can be completed, if it is fun, or if it has problems like \textit{bugs} or \textit{glitches}. 

Usually, large companies (so-called ``AAA studios'') have in-house Quality Assurance (QA) teams that perform playtesting. Smaller companies either use outsourcing or let their developers playtest their games, like most independent developers (indie)\footnote{\url{https://kotaku.com/quality-assured-what-it-s-really-like-to-play-games-fo-1720053842}}. No matter the company, AAA or indie, manual playtesting does not scale to the size of the game. The bigger the game, more playtesters are needed.

The manual, ad-hoc nature of playtesting gives space to improvements in the process. Yet, automated playtesting is not a simple activity. Recently, with the success of Machine Learning (ML) models mastering video games, researchers (and game development companies to a lesser degree) also began to use these models for playtesting games.
Our research question is, thus: \textbf{how suitable are the video game automated testing techniques from the academic literature for the video game developers?}

To answer these questions we conduct this study by (1) performing a \textbf{literature review} on video game automated testing techniques and (2) applying an \textbf{online survey} with video game developers. We asked them to assess some academic solutions on their desirability, viability, and feasibility.

The results of the literature review show a rise in research topics related to automated video game testing in recent years. Yet, most testing tools and frameworks are more concerned with the performance of the ML models instead of the testing objective. The survey results show that game developers are skeptical about using automated agents to test games.

We conclude that there is still a long way to go for video game testing. Especially on how should we test video games. For the practitioners, there is a need for new testing approaches that did not disrupt the developer workflow. As for the researchers, the focus should be on the testing goal and testing oracle. Finally, always offer a replication package and source code.

The paper has the following structure. Section~\ref{sec:backround} discusses software testing and video game testing. Section~\ref{sec:method} explains the method used for the Literature Review and the Developers' Survey. Section~\ref{sec:results} shows the Review and the Survey results. Section~\ref{sec:discussion} discusses the results and threats to their validity. Section~\ref{sec:conclusion} concludes the paper with future works.

\section{Background}
\label{sec:backround}

\begin{figure}[ht]
    \centering
    \includegraphics[width=1\linewidth]{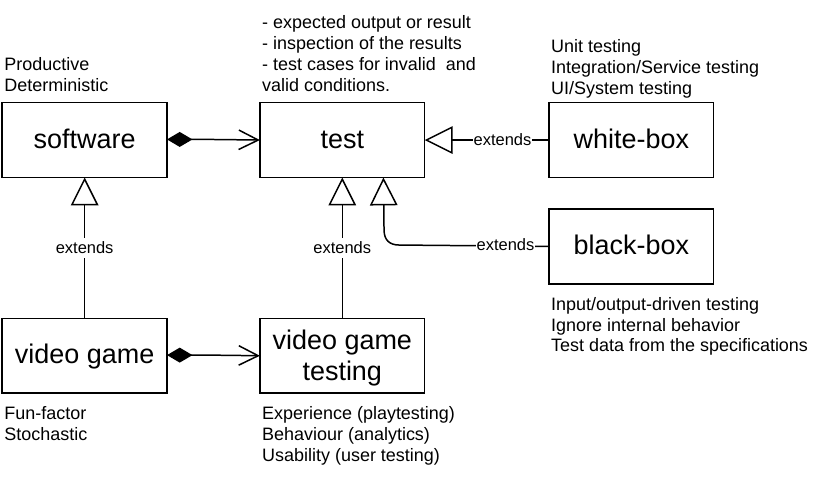}
    \caption{Software Testing \& Video Game Testing summary.}
    \label{fig:testing-uml}
\end{figure}

In this section we describe the concepts of \textit{software testing} and \textit{video game testing}. \autoref{fig:testing-uml} shows the summary of the concepts.
There is no clear definition of what is video game testing and what it implies. This subject requires further investigation and discussion, which is out of the scope for this paper. Thus, for this study, we include technical testing, Quality Assurance (QA), and Game User Research (GUR) under the ``umbrella'' of video game testing. 

\subsection{Software Testing}

Software testing is a discipline that gains more importance as the software industry evolves. Faults and their corrections are among the main factors leading to budget overruns \cite{Durelli2019}. It is not surprising that, nowadays, testing accounts for more than 50\% of the total costs of software development \cite{Garousi2016}.

Software testing is part of the process to verify and improve the quality of a software \cite{Naik2011}, so-called System Under Test (SUT) \cite{Durelli2019}. Testing analysis can be static (based on source code) or dynamic (using SUT executions). The objectives of testing are to check if the SUT works; if not, to find the faults. Fault (or defects) refers to the \emph{cause} of an error, which is the problematic \emph{state} of a SUT that might cause it to not \emph{behave} according to the specification, which may lead to a failure \cite{SWEBOK2014}. 

A test case must consist of two components \cite{Myers2011}:  (1) a description of the \textit{input} data to the program and (2) a precise description of the correct output of the program for that set of input data---an \textit{oracle}. The oracle is used to verify the correctness of the outputs produced by the SUT. It is usually performed by testers but an oracle can be a specification or even another program \cite{Durelli2019}.

Tests are commonly divided into three levels \cite{Naik2011, SWEBOK2014, Aniche2021}: \textit{Unit testing} as the smallest testing part; \textit{Integration testing} for complex integration of classes/procedures; and \textit{System testing} for the main (or risky) flow of the application.
Also, \textit{regression testing} corresponds to a subset of the previous tests to ensure that changes do not break previously-working code while \textit{acceptance testing} is used by clients to assess the final product.

Software tests can be performed by the development team, for which the internals of the SUT are known (white-box testing \cite{Aniche2021}), or by an external party focusing on the SUT functionalities (black-box testing \cite{Mariani2015}) \cite{Naik2011}. Black-box testing only uses the SUT specification to generate and verify test cases, not the internal structure of the SUT.

\subsection{Automated Software Testing}

As the scope of software systems keeps increasing, the testing process also becomes complex. The classic answer of software engineers to reduce cost and complexity is automation \cite{Durelli2019}. Test automation reduces the cost and time used during the testing process, improves efficiency, and reduces human errors \cite{Wiklund2017}. It is ``the use of special software (separate from the SUT) to control the execution of tests and the comparison of actual outcomes with predicted outcomes'' \cite{Garousi2016}.

Testing scripts are the common method of automated software tests. They consist of a pre-defined sequence of actions (as inputs) compared with manually defined oracles \cite{Ricca2021}. 
The same authors also reported the most common automated testing ``solutions'' in research papers. Test generation (test and data) is the main one. Also maintenance of testing scripts and debugging.

\subsection{Video Game Testing}

There is a blurry line between software testing and video game testing (sometimes referred to as ``playtesting''). We choose to consider video games as software with a different purpose: provide an experience (engagement) to the player\footnote{Here we are referring to games that provide only entertainment, not educational or training games.}. Video games, aside from having code, also integrate artistic elements (sound, 2D/3D graphics, narrative, etc.). 
Also, in video games, aside from testing techniques like white-box and black-box testing, developers must assess other attributes, like experience and usability \cite{Politowski2021a}.

Video game testing is a dynamic analysis of the SUT (a version of the game or a Game Under Test), it is usually performed manually by video game testers in playtesting sessions. Besides, to check if the game works and find faults, testers also assess if the game version is fun or engaging (among many other things). Yet, the fun-factor is the main specification of a game: the game, more than anything, must be fun.

Usually, the process to reproduce a problem is described in a report (bug report) written by testers, detailing all the necessary steps to reach the point where the problem occurred. The report consists of text documents with a title, a short description, and the steps \cite{Levy2009}. 
Often, it is hard to replicate the steps (inputs) made by the tester. This situation is even more cumbersome if the game is not deterministic, where the same input does not produce the same output. As randomness is considered a desired feature in games allowing players to keep engaged with the same game (replayability), reproducibility becomes a challenge.


\subsubsection{Game User Research (GUR)}

Game User Research (GUR) is the research field that focus on  \textit{usability} and \textit{user experience} (UX) in video games. This involves any aspect of a video game with which players interact, like menus, audio, artwork, underlying game mechanics, etc. Testing video games involve trying to answer \textit{why} the player is doing something \cite{Drachen2018}. 

In practice, GUR involves many fields, like human-computer interaction, psychology, graphic design, marketing, computer science, analytics, etc. Different than QA and technical game testing, GUR methods focus on evaluating players by observing them interact with the game. The goal is to improve the game using empirical evidence from experimentation and testing \cite{Drachen2018}.

There are different methods to assess the players, which differ across development phases. For example, during pre-production, the main concern is to test the core game loop using the prototypes; while in later phases, the focus is on balancing and tuning. Finally, as for the GUR methods of testing, refer to \cite{Drachen2018}.

There is a difference between GUR and video game testing. Both use playtesting sessions, but, GUR research deals with subjective aspects of the game while video game testing focuses on finding bugs and other technical aspects.

\subsubsection{Automated Video Game Testing}

In a previous work \cite{Politowski2021a}, we briefly discussed automated video game testing issues. According to the academic literature, the main issues are related to:

\begin{itemize}
    \item Coupling: code of game mechanics and UI mixed;
    \item Scope: the game is too big, hard to cover everything;
    \item Randomness: same input, different output;
    \item Changes: core game design changing constantly;
    \item Cost: the engineer is more expensive than a game tester;
    \item Time: programmers focus on creating to fulfil the deadline;
    \item Fun-factor: how to automatically assess the fun?
\end{itemize}

In the same paper, we also investigated the gray literature. According to 22 different game projects, \textit{lack of testing} is the main concern. Moreover, automated testing techniques are in their infancy. For example, some major game development companies are employing bots\footnote{\url{https://www.gdcvault.com/play/1026281/ML-Tutorial-Day-Smart-Bots}} and automatic functional testing using specific game engine features\footnote{\url{https://www.youtube.com/watch?v=KmaGxprTUfI}}.

Finally, test automation in game development is overlooked, as it relies on manual human testers because of all the reasons already presented. Thus, there is a need for new initiatives to find other ways to automate testing games. Ways that do not rely only on human playtesters.

\section{Method}
\label{sec:method}

In the previous section, we discussed what is video game testing. In this section, we present the method in two parts: the literature review, where we search for automated video game testing papers, and the survey with video game developers using an online form, where we ask the respondents to assess the solutions we found in the papers.

\subsection{Academic Literature}

The goal of the literature review is to search, identify, and catalogue automated video game testing techniques presented in academic papers. Instead of starting from scratch, we used a recent study \cite{albaghajatiVideoGameAutomated2020} that already collected works about video game testing. The authors grouped 51 papers according to the approach: search-based, goal-directed, human-like, scenario-based, and model-based.
We decided to extend this work because (1) we found papers about video game testing that were not part of the original dataset, and (2) some of the works in the dataset have solutions too distant from the video game industry reality, making them hard to apply in real-life projects.
Thus, to expand it, we performed full snowballing \cite{Wohlin2014} and further exclusion criteria. 
We used the following criteria for the title and abstract reading:

\begin{itemize}
\item Inclusion criteria:
\begin{itemize}
    \item Paper must be about automated video game testing;
\end{itemize}
\item Exclusion criteria:
\begin{itemize}
    \item Papers about formal or model validation;
    \item Papers about gamification;
    \item Papers about serious games;
    \item Papers about using games for educational purposes;
    \item Papers about using games for medical purposes;
    \item Papers not written in English;
\end{itemize}
\end{itemize}

The final dataset consists of \textbf{166 papers} from 2004 to 2021. Among them, 81 are journal articles, 70 are conference papers, 12 are theses (masters and Ph.D.). There are also 1 report, 1 book, and 1 book chapter. We read all papers, considering five variables to include them or not:

\begin{itemize}
\item \textbf{Study type} (Theoretical/Applied): If the authors produced any practical solution or tool for testing.
\item \textbf{Testing} (True/False): If the paper is about testing games.
\item \textbf{Automated} (True/False): If the testing is somehow automated.
\item \textbf{Machine Learning} (True/False): If the testing uses any machine learning model.
\item \textbf{Test Objective} (String): The goal of the testing, i.e., \textit{balancing} the game, \textit{finding bugs},  etc.
\end{itemize}

From the full reading, we found 114 papers that present some sort of applied approach, that is, a solution or a tool for game testing. Among these, 80 used at least one automated step and \textbf{53} used some type of machine learning model. We use this filter to exclude papers that use manually written scripts. The full list of papers is with the support material at \url{https://doi.org/10.5281/zenodo.5854809}.

\subsection{Developer Survey}

We divided the survey into four sections. The first section discussed the respondent \textit{background}; the second asked them about their \textit{manual playtesting} activities when developing a game; the third asked the participants to assess academic techniques/solutions for \textit{automated playtesting}; and, the fourth contained optional open questions about the \textit{future of game testing}. The questions, as well as the answers, are available at \url{https://doi.org/10.5281/zenodo.5854809}.

\subsubsection{Preparing the Questions}

To reduce the scope of the survey, we focused on the three most common testing objectives: \textit{balancing}, \textit{exploration}, and \textit{finding bugs}. To choose which paper (solution) we put on the survey, we considered four aspects (see below). \autoref{tab:papers-survey} shows the papers selected to be in the survey.

\begin{itemize}
\item Include papers that not only propose a solution but actually implemented it;
\item Avoid papers that use frameworks or platforms that are deprecated;
\item Include papers where the authors validated (or evaluated) their solution;
\item Include papers that provide source code (replication package).
\end{itemize}

\begin{table}[ht]
\caption{Selected papers (solutions) we asked participants of the survey to assess. Except \citet{ariyurekPlaytestingWhatPersonas2021}, the other six papers are also discussed by \citet{albaghajatiVideoGameAutomated2020}.}
\label{tab:papers-survey}
\begin{tabular}{@{}lllll@{}}
\toprule
\# & Author & Test Obj. & Game tested \\ \midrule
1 & \citet{gudmundssonHumanLikePlaytestingDeep2018} & Balancing     & Match 3 \\
2 & \citet{roohiPredictingGameDifficulty2020} & Balancing     & Puzzle \\ \addlinespace
3 & \citet{gordilloImprovingPlaytestingCoverage2021} & Exploration     & 3D third person \\
4 & \citet{ariyurekPlaytestingWhatPersonas2021} & Exploration     & Doom \\
5 & \citet{zhengWujiAutomaticOnline2019} & Exploration     & MMOG \\ \addlinespace
6 & \citet{pfauAutomatedGameTesting2017} & Finding bugs     & Adventure \\
7 & \citet{ariyurekEnhancingMonteCarlo2020} & Finding bugs     & 2D Adventure \\ \bottomrule
\end{tabular}
\end{table}

\subsubsection{Putting the papers' solution in the survey}

To explain the paper for the survey participants, we divide the papers' solutions into four parts, with a short description for each: the \textit{GOAL} of the paper; \textit{HOW} the authors accomplished that goal (the method); the role of \textit{AUTOMATION} in the solution; and the final \textit{RESULTS} of the paper. For example, for Paper \#3 \citet{gordilloImprovingPlaytestingCoverage2021}, we wrote:

\begin{quote}
``GOAL: Testing coverage in complex 3D environments. HOW: Fully traverse the environments using autonomous agents (bots). AUTOMATION: Automate the collection of some playtest data. RESULTS: The agent can reach areas that should not be inaccessible.''    
\end{quote}

For each question, we asked the participants to assess (a) \textbf{Desirability} -- Is this something you would like to use for game testing? (b) \textbf{Viability} -- Do you think it can be implemented in your workflow? (c) \textbf{Feasibility} -- For you, do you believe this idea would bring benefits to game testing?

\subsubsection{Sending the Survey}

We sent the survey to online communities, groups, and forums for game developers and game testers, including on \textit{Reddit}, \textit{LinkedIn}, \textit{Facebook}, \textit{Discord}, and \textit{Itch.io}.

\section{Results}
\label{sec:results}

\subsection{Literature Review}

From our dataset of 166 papers, 80 of them suggested some types of applied solutions for video game testing. These solutions were either fully-automated (as mentioned by their authors) or semi-automated. Finally, 53 papers used some types of machine learning models to train agents and playtest games.

In that set of 53 papers, the most discussed test objectives were \textit{balancing the gameplay} (19 papers). The remaining ones were \textit{game exploration} (11 papers), \textit{finding bugs} (6 papers), and \textit{player modeling} (6 papers). Also, testing the \textit{game mechanics}, \textit{UI}, \textit{UX}, \textit{visual correctness}, \textit{collision}, and \textit{visualization}.

\begin{figure}[ht]
    \centering
    \includegraphics[width=1\linewidth]{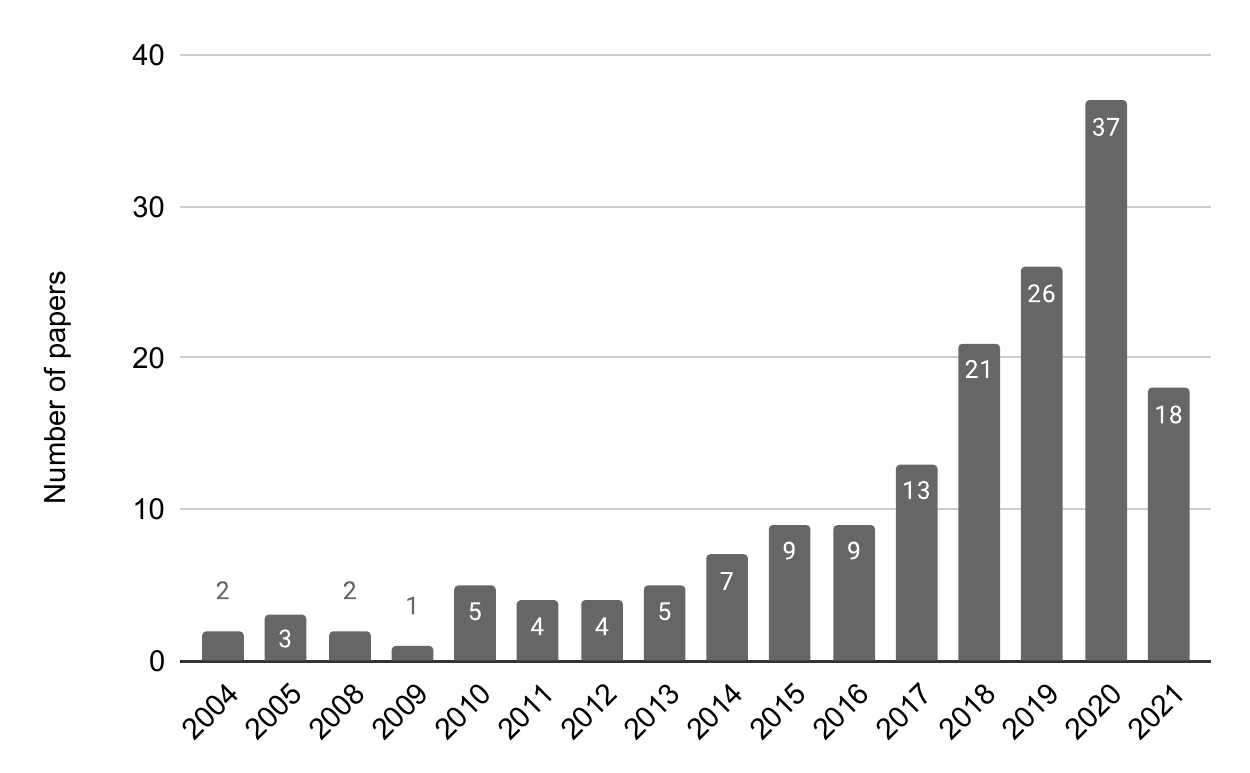}
    \caption{Histogram of all the 166 papers. There is clear rise of the subject ``automated video game testing'' in recent years.}
    \label{fig:chart-year}
\end{figure}

According to \autoref{fig:chart-year}, automated video game testing is an emergent field. Even with a drastic decrease in 2021\footnote{The study was made in November 2021.}. The majority of the studies were from the last 2-3 years, especially 2020 with 37 papers.

The academic papers that used ML focused on the model instead of the game testing problem. Even filtering the papers that offer solutions and evaluations, their applicability in real-world scenarios does not seem viable. We identified some main issues about the papers:

\begin{itemize}
\item The tools (game engines, games, etc.) used in the experiment are too simple, incomplete, or academic projects.
\item The testing objectives are not clearly defined, sometimes with phrases like ``it can be used to test the game''.
\item There is no oracle or it is made manually \emph{after} the autonomous agents play the games.
\item The source code (replication packages) are often not available, which makes it impossible to replicate or re-use the proposed solutions.
\end{itemize}

\subsection{Survey}

We had a total of 12 accepted responses. The majority (58\%) of the respondents have more than four years of experience in game development. The same proportion reported working full-time in a game company. Almost all respondents have different roles, from software tester to game designer. All play video games as a hobby and the great majority (92\%) have experience developing traditional software.

All respondents use manual playtesting regularly. Some of them test the game within the game engine. None of them use scripts to playtest games. Their testing objectives include \textit{searching for bugs} followed by \textit{exploring the game content} and \textit{balancing the game mechanics}. Finally, \textit{crash} and \textit{stuck} behaviour is what the developers usually try to spot when testing followed by \textit{graphical}, \textit{collision}, and \textit{performance} issues.

\begin{table*}[ht]
\caption{Survey results related to the solutions (paper's ideas).}
\label{tab:survey}
\begin{tabular}{@{}lllccc|ccc|ccc@{}}
\toprule
\# & Solution & Test Obj. & \multicolumn{3}{c|}{Desirability} & \multicolumn{3}{c|}{Viability} & \multicolumn{3}{c}{Feasibility} \\ \cmidrule(l){4-12} 
 & & & Yes & Not Sure & No & Yes & Not Sure & No & Yes & Not Sure & No \\ \midrule
1 & \citet{gudmundssonHumanLikePlaytestingDeep2018} & Balancing & 6 & 5 & 1 & 2 & 5 & 5 & 5 & 3 & 4 \\
2 & \citet{roohiPredictingGameDifficulty2020} & Balancing & 3 & 6 & 3 & 1 & 8 & 3 & 1 & 6 & 5 \\ \addlinespace
3 & \citet{gordilloImprovingPlaytestingCoverage2021} & Exploration & 8 & 2 & 2 & 4 & 4 & 4 & 6 & 2 & 4 \\
4 & \citet{ariyurekPlaytestingWhatPersonas2021} & Exploration & 3 & 7 & 2 & 4 & 4 & 4 & 3 & 4 & 5 \\
5 & \citet{zhengWujiAutomaticOnline2019} & Exploration & 5 & 5 & 2 & 2 & 7 & 3 & 4 & 4 & 4 \\ \addlinespace
6 & \citet{pfauAutomatedGameTesting2017} & Bugs & 6 & 4 & 2 & 5 & 6 & 1 & 7 & 2 & 3 \\
7 & \citet{ariyurekEnhancingMonteCarlo2020} & Bugs & 5 & 5 & 2 & 2 & 7 & 3 & 6 & 3 & 3 \\ \bottomrule
\end{tabular}
\end{table*}


\subsubsection{Solution \#1}

The solution \#1 by \citet{gudmundssonHumanLikePlaytestingDeep2018}, titled ``Human-Like Playtesting with Deep Learning'', uses autonomous agents to predict the difficulty of a new game level automatically. 
To train the agents the authors used Convolutional Neural Networks (CNN) in a grid-like structure (a Match-3 game called Candy Crush), using a discrete actions space. The same method used by AlphaGO\footnote{\url{https://deepmind.com/research/case-studies/alphago-the-story-so-far}}. \autoref{tab:survey} shows that this solution is desired and feasible for the respondents. Yet, it is not viable according to them. Among the reasons are the necessity of ``lots'' of data and the need for building the testing pipeline from scratch. Another problem is the time needed to train the agents. They also mentioned that Match-3 games are not as random as they seem: designers deliberately make choices to avoid players getting stuck.


\subsubsection{Solution \#2}

The solution \#2 by \citet{roohiPredictingGameDifficulty2020}, titled ``Predicting Game Difficulty and Churn Without Players'', uses gameplay data from autonomous agents and human playtesters to check the pass rate of new levels automatically. To train the agents, the authors used Deep Reinforcement Learning with the Proximal Policy Optimization (PPO). They tested a puzzle game using Unity ML-agents\footnote{\url{https://unity.com/products/machine-learning-agents}}. Respondents were more averse to this idea compared to the previous solution \#1.
Because the solution used players' churn data, they mentioned that it is hard to spot precisely why (and where) players abandon games. One respondent stated: ``It's also something that has more value when captured during soft launch without much effort.''.



\subsubsection{Solution \#3}

The solution \#3 by \citet{gordilloImprovingPlaytestingCoverage2021}, titled ``Improving Playtesting Coverage via Curiosity Driven Reinforcement Learning Agents '', uses autonomous agents to fully traverse complex 3D environments. To train the agents, the authors used Reinforcement Learning with the PPO algorithm. They tested a complex 3D environment using an undisclosed game engine. Compared to all others solutions, this was the most desired, as it deals with a situation that is lengthy to test manually. Some of the answers agreed with the approach for the exploration of edge cases, allowing ``more obvious glitches be caught during manual testing''.


\subsubsection{Solution \#4}

The solution \#4 by \citet{ariyurekPlaytestingWhatPersonas2021}, titled ``Playtesting: What is Beyond Personas'' uses autonomous agents with different personas (killers, explorers, etc.) to discover different paths at the level. To train the agents, the authors used Reinforcement Learning with the PPO  algorithm. They used the General Video Game Artificial Intelligence (GVG-AI)\footnote{\url{https://gaigresearch.github.io/gvgaibook/}} and VizDoom\footnote{\url{http://vizdoom.cs.put.edu.pl/}} frameworks to evaluated their solution. The respondents were not sure if they wanted this solution and were divided about the viability and did not think it is applicable in practice.


\subsubsection{Solution \#5}

The solution \#5 by \citet{zhengWujiAutomaticOnline2019}, titled ``Wuji: Automatic Online Combat Game Testing Using Evolutionary Deep Reinforcement Learning'', uses autonomous agents (bots) with different goals to explore game states and corner cases. The authors developed a testing agent called ``Wuji'', which uses Evolutionary Algorithms and multi-objective optimization to explore game space. They used Reinforcement Learning to direct the agent while exploring the state space. They evaluated the solution using an undisclosed MMOG. The respondents are not sure about the viability of this solution and are divided about its feasibility.



\subsubsection{Solution \#6}

The paper \#6 by \citet{pfauAutomatedGameTesting2017}, titled ``Automated Game Testing with ICARUS: Intelligent Completion of Adventure Riddles via Unsupervised Solving'', uses autonomous agents to complete the game like a ``speedrun''\footnote{A speedrunner aims to complete a game as quickly as possible.} and spot crashes/freezes and blocker (soft lock). To automate the process, the authors used the script language Lua on top of the Visionary game engine\footnote{\url{https://www.visionaire-studio.net/}}. According to \autoref{tab:survey}, this paper idea is the most feasible of all solutions and very desirable and viable.

\begin{quote}
\textit{``The biggest bang for the buck would be as a build acceptance test on a CI/CD pipeline, making sure the catch obvious blocking bugs. Otherwise it drops significantly in usefulness''}. -- survey respondent about paper \#6 \cite{pfauAutomatedGameTesting2017}.
\end{quote}


\subsubsection{Solution \#7}

The solution \#7 by \citet{ariyurekEnhancingMonteCarlo2020}, titled ``Enhancing the Monte Carlo Tree Search Algorithm for Video Game Testing'', uses agents to generate sequences that can be replayed, to explore games and spot bugs. The authors modified the Monte Carlo Tree Search policy to use different strategies. They also used the  General Video Game Artificial Intelligence (GVG-AI) in a 2D adventure game. The respondents reported viability as a problem. The respondents' concern is that the authors used pre-defined bugs: one respondent stated that ``You don't want to find the bugs you already know about''. 


\subsection{Future of Video Game Testing}

We also asked the game developers open questions about the future of video game testing. First, we asked \textit{``What is the most important aspect of video game testing?''}. The answers varied:

\begin{itemize}
    \item Identifying areas for improvement in the game;
    \item Helping make the game work as the players expect;
    \item Making sure everything works is secondary;
    \item When playing a game, it must feel right;
    \item Matching specifications (game design requirements) is not enough;
    \item Testing to check how players perceive the game;
    \item Testing to check the UX.
\end{itemize}

We also asked what could help game testers do their jobs: \textit{``Currently, what is lacking in the video game industry that could help video game testers do their jobs?''}. Respondents mentioned easy-to-maintain test automation that is decoupled from the game under test. The lack of testing process and lack of engineer expertise were also mentioned.

\begin{quote}
\textit{``I think game testers would greatly benefit from learning standard software testing and engineering from the rest of the industry. A lot of the testing is done manually by non technical people with no knowledge about game engines, backend services, graphics API and so on. This also applies to developers. While they are good at making games, they are terrible at engineering, don't follow good practices. I have never seen an unit test written in a game, for instance. It would also help if they were actually agile instead of doing waterfall cycles of months if not year.''} -- survey respondent about what could help game testers do their jobs.
\end{quote}


When we asked \textit{``In 10 years, how do you think video games will be tested?''}, the respondents were skeptical. They believe the majority of game companies will still be using manual testing. Engineers will still be working mainly on the games instead of building testing tools.

\section{Discussion}
\label{sec:discussion}

\subsection{Academic Literature}

The recent rise of papers about automated testing in video games shows a trend in software engineering testing. We identified four main reasons: (1) the rise of machine learning models and solutions applied to every aspect of software development; (2) the achievements of machine learning models on playing video games; (3) all the new tools for writing and deploying machine learning models; and, (4) the need to automate testing games, as the scope, cost, and complexity of games keep on increasing.

Yet, only half of the studies we found offer some kind of applied solution for automated testing. Among them, we perceived a lack of focus on the real problem: the testing part.
Few papers present testing oracles or even discuss them, relying on manual assessment.
The testing goals are not clear enough, sometimes it is even hard to find in the paper what exactly the authors were trying to test. Also, the lack of replication packages makes the solutions hard to evaluate and reuse. 

Finally, there is a segmentation of the video game testing papers. Our dataset contains papers that pertain to Artificial Intelligence, Computer Science, Software Engineering, and Video Game Design. These different fields have different goals. For example, AI papers focus on creating the models, while those on Software Engineering focus on testing concerns, like the oracles. A proper solution for automated video game testing needs all these fields of study working together. A new field of study and--or specific venues for discussing this matter need to arise.

\subsection{Developers' Survey}

In general, the respondents were skeptical about the solutions of the academic papers. The solutions with a more straightforward process were better received, as the Solution \#3 by \citet{gordilloImprovingPlaytestingCoverage2021}. Training agents to test games is seen as a ``waste of time and money that can be spent somewhere else'', which is a recurrent narrative in the video game industry. Building a complex testing pipeline requires dedicated software engineers, which cost more than manual testers. For small games (indie developers, for example) none of the presented seven solutions are suitable.

A respondents' recurrent concern is the up-front cost of building testing tools and training agents. There is a clear need for open-source, general tools that allow video game developers to test their games. These tools must work with a wide range of game types, without lengthy, complex customization. They should work directly with game engines, which are the main tools for all video game developers. 
For example, the Unreal engine (version 4) has built-in functionality tests. Yet, to test a feature requires doubling the developer's effort, like traditional unit tests. Game developers need an easier way to automate tests in their games.

Some respondents were concerned about AI replacing game testers. We believe that this is not going to happen soon. Even if better machine learning models could be built, spotting inconsistencies, odd behaviours, and verifying the ``fun'' in games, need humans, who can do it trivially. 

\subsection{Weaknesses of the Study}

It is not trivial to translate complex solutions from academic papers to practical solutions for game developers. We chose to keep the description of the solutions as simple as possible, hiding technical details about the machine learning models and focusing on the testing objectives. 

Testers do not have a complete vision of the development process, as they are assigned specific tasks. They might have difficulties assessing solutions that often disrupt their normal workflows. We expanded the survey's scope to include developers, designers, and managers.

The low number of respondents might introduce bias to the study. We tried to create a simple way to explain complex ideas from the papers to developers. Even hiding the complexity, completing the form took more time and effort from the developers than we expected. This could explain why the low number of answers.

We did not ask the respondents the type of studio they work in. If it is a big, medium, small, or even an ``one-man-army'' independent developer. This also can contribute to the bias in the results.

Finally, while the respondents may be professional game developers, they likely do not have the AI knowledge to make an accurate assessment of the utility of the solutions. To mitigate that, we tried to hide the complexity of the solutions.

\section{Conclusion}
\label{sec:conclusion}

This paper is an exploratory study that investigated the gap between the solutions for automated video game testing in the academic literature and video game developers' needs in practice. We performed a literature review on automated video game testing with 166 papers. Among these papers, we identified 53 solutions for automated video game testing. After filtering the most promising solutions, we performed an online survey with video game developers. We asked respondents to assess seven solutions about their desirability, viability, and feasibility. 

The results of the literature review show a rise in research topics related to automated video game testing in recent years. Yet, most testing tools and frameworks are more concerned with the performance of the ML models instead of the testing objective. The survey results show that game developers are skeptical about using automated agents to test games.

We conclude that there is still a long way to go for video game testing. Especially on how should we test video games. For the practitioners, there is a need for new testing approaches that did not disrupt the developer workflow. As for the researchers, the focus should be on the testing goal and testing oracle. Finally, always offer a replication package and source code.

\paragraph{Limitations and Future Works:} In this study we tried to check and discuss academic solutions from the point of view of developers. Yet, the low response rate made it impossible to reach any concrete conclusion. Thus we still believe this work can serve as a pilot study. In future works, we plan to gather more respondents to the survey and expand it using interviews.

\begin{acks}
The authors were partly supported by the NSERC Discovery Grant and Canada Research Chairs programs.
\end{acks}

\bibliographystyle{ACM-Reference-Format}
\bibliography{main.bib}
\end{document}